\documentclass[%
 preprint,
 amsmath,amssymb,
 aps,
]{revtex4-2}
\usepackage{url}
\usepackage{hyperref}
\usepackage{lipsum}
\usepackage{graphicx}
\usepackage{dcolumn}
\usepackage{bm}

\makeatletter
\newcommand\emailx[1]{%
\move@AF%
\def\@affil{{\normalfont\,#1\strut}{}}%
}%
\makeatother

\begin{document}

\title{Emergence of synchronised rotations in dense active matter with disorder}

\author{Danial Vahabli$^{1,2}$ \& Tamas Vicsek$^{3,4}$}

\address{$^1$ Department of Physics, Middle East Technical University, Ankara 06800, Turkey}
\address{$^2$ Institute for Advanced Computational Science, Stony Brook University, Stony
Brook, NY, United States of America}

\address{$^3$ Department of Biological Physics, E\"{o}tv\"{o}s University, P\'azm\'any P\'eter s\'et\'any 1A, H-1117, Budapest, Hungary}

\address{$^4$ MTA-ELTE Statistical and Biological Physics Research Group of HAS, P\'azm\'any P\'eter s\'et\'any 1A, H-1117, Budapest, Hungary}

\emailx{vicsek@hal.elte.hu}

\vspace{10pt}

\bigskip

{\let\newpage\relax\maketitle}

\section*{Abstract}

We consider the rich variety of collective motion patterns emerging when aligning active particles move in the presence of randomly distributed obstacles - representing quenched noise in two dimensions. In order to get insight into the involved complex flows and the transitions between them we use a simple model allowing  the observation and analysis of behaviours that are less straightforwardly  accessible by experiments or analytic calculations. We find a series of symmetry braking states in spite of the applied disorder being isotropic. In particular, as the level of perturbations is increased, the system of self-propelled particles changes its collective motion patterns from i) directed  flow ii) through a mixed state of locally directed or locally rotating flow to iii) a novel, globally synchronized rotating state thereby the system violating overall chiral symmetry. Finally, this phase crosses over to a state in which iv) clusters of locally synchronized rotations are observed. The way of change from polar flow to overall synchronization can be interpreted as indicating a non-reciprocal phase transition. Our simulations suggest that, when both present, quenched rather than shot noise dominates the behaviours.

\section*{Introduction}

Systems made of many self-propelled units exhibit a broad range of fascinating collective motion patterns, the examples ranging from macro-molecular level to groups of organisms \cite{Stumpter2010,Vicsek2012,Marchetti2013,Cavagna2014,Alert2022} and the observed behaviours are reminiscent of those in equilibrium many particle systems as well as are manifestations of new kinds of features. While a phase change from a disordered to ordered state occurs both in equilibrium and far from equilibrium, a conspicuous new phenomenon in systems of actively moving unis is the emergence of several kinds of rotational motions due to various origins. The very different types of systems, in which units move along circular trajectories, that have been studied include a wide range of complexity from molecular motors-driven biological macromolecules \cite{Schaller2010,Sumino2012}, colonies of bacteria, \cite{Czirok1996,Yamamoto2021}, cells \cite{Mehes2014,Battersby2015,Be'er2019,Hong2021}, ants \cite{Couzin2003a}, fish \cite{Calovi2014} and even groups of people \cite{Kolivand2021}. 

In these works, it was implied that the reason behind the rotating patterns of group motion was either a global spatial confinement (which can also be due to the entities preferring to stay close) or a violation of the left-right symmetry on the individual level (see, e.g., bending of a macromolecule, chirality-broken swimming of sperm cells \cite{Riedel2005}), spiral background director \cite{Koizumi2020}). Even the flexible, fluctuating boundary of a droplet can result in a single drop-spanning vortex 
\cite{Kokot2022} of active particles.
An important paradigm of the widely observed systems of self-propelled entities is soft active matter \cite{Toner2005,Marchetti2013,Chate2020} consisting of particles interacting through alignment forces and being subject to fluctuations while moving with approximately the same speed. Experiments (allowing the control of the conditions) in which rotations have been observed are typically carried out using macromolecule essays \cite{Schaller2010,Sumino2012}, colonies of bacteria \cite{Ben-Jacob2000,Watanabe2002, Wakita2015,Reinken2020} or, since the introduction of the elegant experiments first involving Quincke particles \cite{Bicard2013} (which are colloidal microrollers propelled by an electric field - in the subcritical version of the experiment free standing vortexes were recently observed and modelled, see \cite{Chate2021b}) and finally, in a newer setup, magnetic beads driven by an oscillating magnetic field \cite{Kaiser2017,Han2021}. 
Until very recently the perturbations considered have predominantly been assumed to be of shot noise like, i.e., being uncorrelated in both space and time. In the context of the above, it is natural to ask: what are the main factors leading to rotational motion of active matter? So far such patterns of motion have mostly been observed in situations where the following three conditions had an essential influence: i) a global confinement was present - representing reflective type boundary conditions, ii) a relatively long range (either in time or space) cohesive force kept the self-propelled units within a slowly gliding area (within which the particles moved relatively fast) and, finally iii) there was an intrinsic chirality breaking interaction present (such as, e.g, magnetic force \cite{Han2021}) or a combination of these. 
On the other hand, the units of active matter (macromolecules, bacteria, colloids, etc) are in many cases moving in a disordered environment with inhomogeneities representing temporally (and/or spatially) correlated perturbations. Thus, we address the question: can isotropic disorder (quenched noise) result in rotational patterns of motion?

Introducing randomly placed obstacles (quenched disorder) into active matter systems has a relatively short history with early works indicating that particles can get trapped, resulting in genuine subdiffusion and isolated rotation \cite{Chepizhko2013} particles may rotate around obstacles (in a region surrounding a given large obstacle)  collectively \cite{Mokhtari2017}, travelling bands can survive disorder if the interaction is topological type \cite{Rahmani2021}, while swirling patterns were observed in simulations of two kinds of active particles (in active nematics) in \cite{Sampat2021}. Rich - and relevant from the point of our paper - motion patterns were observed in the beautiful experiments on Quincke rollers moving among randomly placed obstacles \cite{Morin2017,Chardac2021}. Flocking in an inhomogeneous medium was studied in \cite{Toner2018} using a sophisticated continuum equation approach and it was shown that under the assumptions they made collective effects were more robust in such an active system than its equilibrium counterpart. On the other hand, randomly placed rotators were shown to destroy global flow beyond a critical density \cite{Das2018}. Quite remarkably, experiments involving both active units and obstacles have mostly been carried out by introducing microscopic obstacles into colonies of swimming bacteria \cite{Galajda2007,Reiken2020}, but these obstacles were arranged according to specific regular patterns. To the best of our knowledge the papers by Nishiguchi et al \cite{Nishiguchi2018} and Reiken at al. \cite{Reiken2020}, reporting on the collective motion patterns of swimming bacteria in the presence of obstacles arranged as the nodes of regular lattices (square, hexagonal and Kagome-like) are the only publication which present experimental evidence of local rotation of living active matter due to inhomogeneity. Although obstacles typically cause an overall slow down of flocking velocity, the situation is more complex and the possible range of phenomena has not still been fully explored. For example, paradoxically - as was shown recently by Kamdar et al. \cite{Kamdar2022} - obstacles can even result in enhanced mobility of bacteria when they move in a complex fluid containing colloidal particles.

Among the numerous interesting questions related to active matter systems one can address the one which is concerned with the nature of the transitions between the observed phases. For example: Is it analogous to those taking place in equilibrium systems? Can it be interpreted in terms of order parameters changing abruptly (if first order type) or continuously (second order phase transition)? Or, alternatively the situation is more complex and in the transitional region new type of behaviours can be observed as a recent theory predicts (Ref. \cite{Fruchart2021}). Since our system is made of 2 kinds of units interacting in an asymmetric manner (the moving ones are repelled by the obstacles, while the obstacles do not react to the moving ones) we may see a behaviour that is analogous to non-reciprocal phase transitions \cite{Fruchart2021} allowing an extended region of mixed moving phases including rotations. Such mixed moving phases (obviously non-existent in equilibrium systems) were observed experimentally and interpreted by simulations \cite{Huber2018} of actomyosin motility assay.

\newpage

\begin{figure}
\centering
\includegraphics[scale = 0.9]{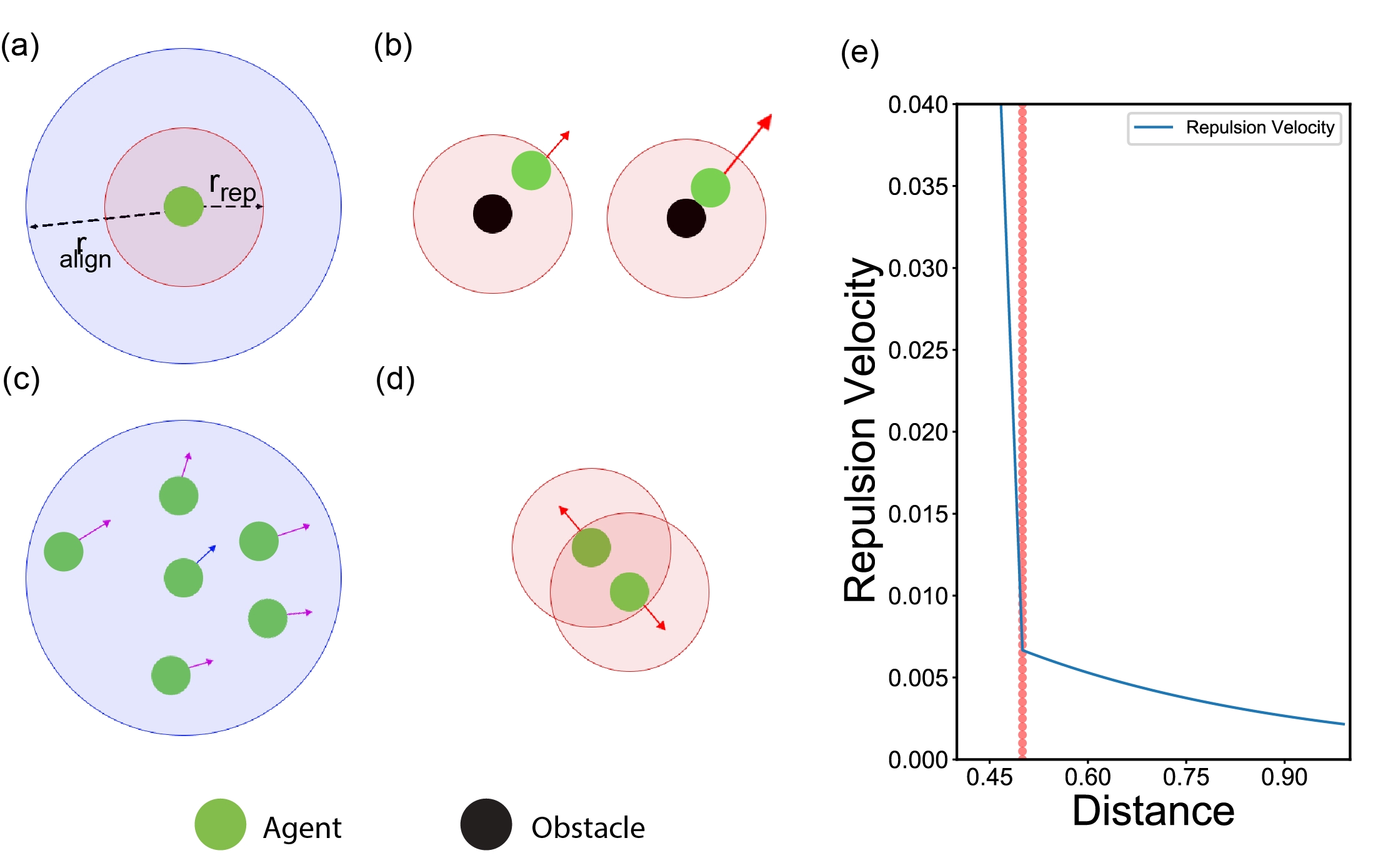}
\caption{Visualisations of the interactions of our soft active particles model. Green circles denote particles and black circles denote obstacles (a) overall view. (b) particle-obstacle repulsion and (c) visualizes particle-particle alignment and (d) particle-particle repulsion. The repulsion terms correspond to a sum of a hard core and an exponentially decaying soft term.  (e) illustrates the magnitude of repulsion}
\label{Model}
\end{figure} 

\section*{Model and definitions}
\subsection*{The rules of our model}

We model the dynamics of self-propelled (active) particles interacting with each other and with randomly distributed obstacles in two dimensions. The main features of the related interactions are visualized in Fig. \ref{Model}. The active particles tend to align their direction of motion if they are within the alignment radius and are repelled by a distance dependent force both by each other and by the obstacles. Motivations for considering soft repulsion have both computational and physical origins (e.g., pure hard core may result in undesirable sudden jumps while hydrodynamic forces between Quincke particles were approximated by a soft potential in \cite{Chate2021b}). The model we are studying can be considered as made of two kinds of particles, A and B, one kind of them moving (A) while the other ones(B) are being fixed in space with the repulsion force acting between all of the particles being identical. Thus, particles B can "push" particles A away, but A cannot change the position of B particles and in this way the interactions are such that particles B represent quenched noise and, at the same time, the effects of repulsion are not reciprocal \cite{Fruchart2021}. \\
The position of each particle is updated at each time step as:

\begin{equation}
\mathbf{v}_i(t+1) = \mathbf{v}_i^{align}(t+1) + \mathbf{v}_i^{rep}(t+1)
\label{V_sum}
\end{equation}
\begin{equation}
\mathbf{x}_i(t+1)= \mathbf{x}_i(t) + \mathbf{v}_i(t+1) \Delta t 
\label{next_pos}
\end{equation}
where $x_i$ and $v_i$ respectively denote the position and the velocity of the  particle and $ \Delta t = 1 $ is assumed in our simulations. Thus, the velocities of the particles are obtained from a combination of an alignment ($\mathbf{v}_i^{align}$) and a repulsion ($\mathbf{v}_i^{rep}$) term which are calculated from the expressions given below.

The alignment term, in the spirit of ref. \cite{Vicsek1995}, has an alignment velocity magnitude $v_{align}$ and a unit vector $\mathbf{e}_i$ pointing in the direction of $\theta_i$:

\begin{equation}
\mathbf{v}^{align}_i(t+1) = v_{align}  \mathbf{e}_i(t+1)
\label{v_align_t+1}
\end{equation}
\begin{equation}
\theta_i (t+1) = <\theta(t)>_i +  \eta_i (t)
\label{theta_i_sum}
\end{equation}
with
\begin{equation}
<\theta(t)>_i = \arctan{\frac{\sum_{j=1} \sin{\theta_j(t) }}{\sum_{j=1} \cos{\theta_j(t) }}}   
\label{theta_i}
\end{equation}
where $j$ runs over the indices of particles in within the alignment radius of particle $i$ including itself, see Fig. \ref{Model}c. In Eq. \ref{theta_i}$ <\theta(t)>_i$ denotes the average direction of the particles being within the alignment radius ($r_{align}$) of particle $i$.  $\eta$ denotes the shot noise which is drawn from a uniform distribution in the interval $[-\frac{\eta}{2},\frac{\eta}{2}]$. \\

When defining the repulsion term we implement rules in the spirit of simulations aiming at interpreting the collective motion of entities having a hard core and an exponentially decaying soft repelling force (see, e.g., \cite{Helbing2000,Chate2021b}). The  particle-particle  and obstacle-particle repulsion terms are given by:

\begin{equation}
\begin{split}
\mathbf{v}^{rep}_i(t+1) = c_{rep}   \sum_{k=1}  (\exp{\frac{r_{i} + r_{k} - \|\mathbf{x}_i(t) - \mathbf{x}_k(t)\|}{A}} + 
\\
+ c_{core} g(r_{i} + r_{k} - \|\mathbf{x}_i(t) - \mathbf{x}_k(t)\|))\mathbf{e}_{ki}(t)
\end{split}
\label{Repulsion}
\end{equation}
\\
 here $k$ runs over the particles and obstacles that are within the repulsion radius ($r_{rep}$) of  particle $i$ excluding $i$. It is important to note here that the velocity and the force terms are used in our approach interchangeably, since we consider - as it is common in active matter studies - the over-damped limit of the dynamics. The constant $c_{rep}$ is the repulsion coefficient which controls the magnitude of the interaction. The first term in the sum is the soft interaction where $r_{i}$ is the core radius of particle and $r_k$ is the core of either the neighbouring particle or obstacle, both the $r_{i}$ and $r_{k}$ are usually equal to $r_{rep}$.  The second term in the sum models the hard core short-range interactions where the particle-particle pair or particle-obstacle pair are touching, here the parameter $c_{core} $ controls the relative strength of the interaction and the $g(x)$ function is 0 if $|\mathbf{x}_i - \mathbf{x}_j| \geq r_k + r_i$ and is equal to its argument otherwise. (see Fig. \ref{Model} b,c,e) Lastly, the $\mathbf{e}_{ki}(t)$ term is the unit vector from particle $k$ to particle $i$. Further, if the value of the repulsion term is less than 1/10 of $c_{rep}  \exp{\frac{r_{i} + r_{k}}{A}}$ we drop its effect. The value of parameter A controls this cut-off radius ($r_{cut-off}$). \\

\subsection*{Model versus recent experiment}

 \begin{figure}
\centering
\includegraphics[scale = 0.7]{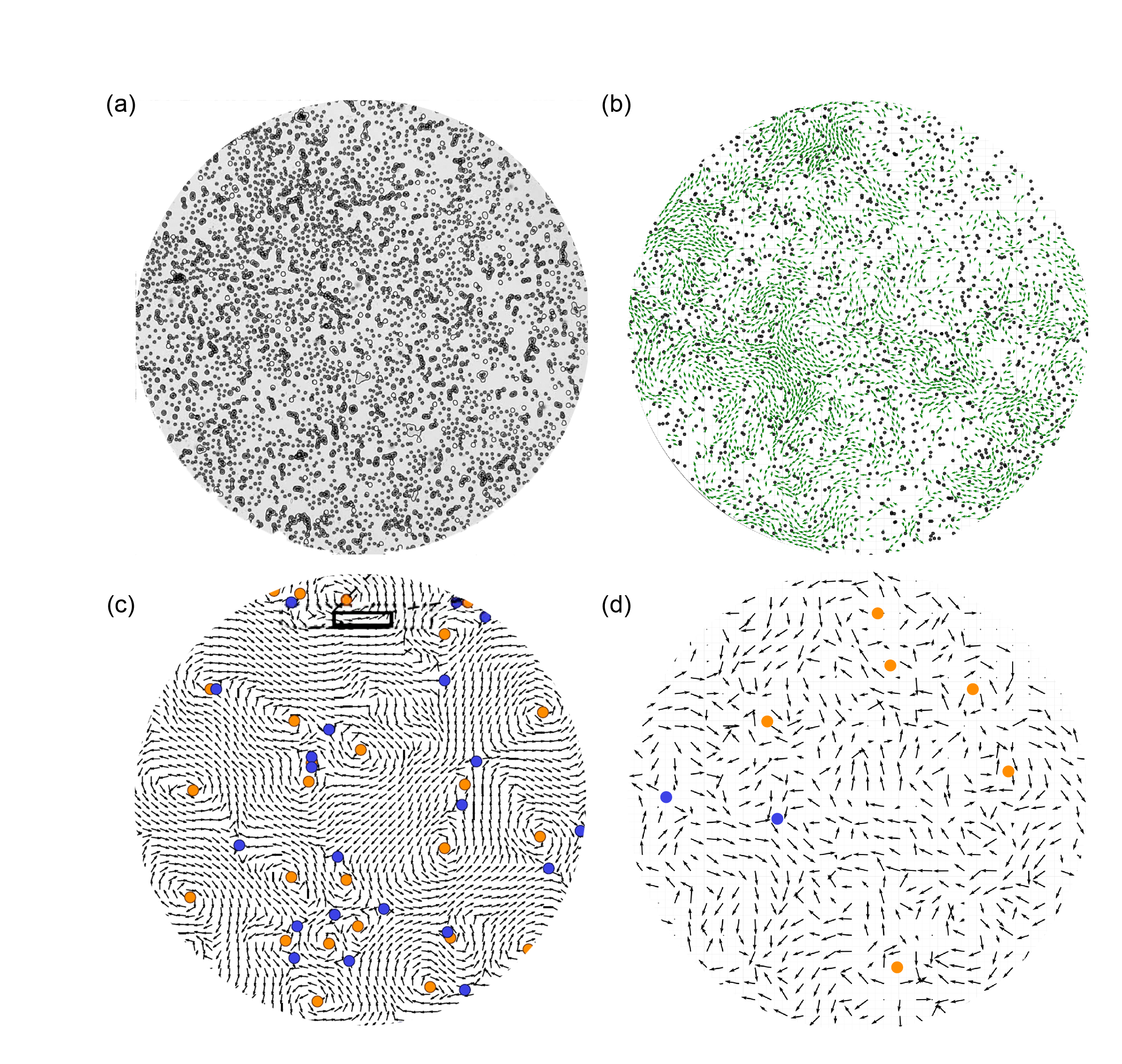}
\caption{Comparison of our results with a figures from  Ref. \cite{Chardac2021}. Figures (a) and (b) show frames from the videos of the motion and figures (c) and (d) show the corresponding velocity fields, where (a) and (c) are from Ref. \cite{Chardac2021} and (b) and (d) were obtained during our simulations. In figure (b)  the green dots are particles and the black ones are the obstacles. The blue and yellow circles in (c) and (d) denote the topological defects in the velocity field. This comparison demonstrates that our model is able to reproduce the "meander" state of motion discovered experimentally by Chardac et.al., Ref. \cite{Chardac2021}.}
\label{PNAS}
\end{figure} 

The above model, in spite of its simplicity, allows us to simulate a rich variety of possible behaviours in systems with active particles moving in a disordered environment. In order to demonstrate this feature of our model we use as reference the very recent results of Chardac et.al \cite{Chardac2021} on the states of active colloidal particles (Quincke rollers) in chambers with randomly distributed obstacles. We find that our active particles exhibit very similar motion patterns to those observed experimentally for an appropriately selected set of the parameters. If the densities (of the moving particles and the obstacles) and their respective sizes, as well as the magnitude of the alignment and repulsive forces are tuned to reflect those in the experiments, we observe an almost perfect matching of the experimental and simulational results. To support this statement we display the two kinds visual data in Fig. \ref{PNAS}. In addition, we also visualized the similarity of the behaviours by comparing the corresponding videos (see: experimental video by \cite{Chardac2021} and our Supplementary video 1). 

\subsection*{Definitions of the order parameters}

To quantify the behaviour of our system, we have defined four order parameters: global flow, average direction, rotational ratio, and global rotation. These parameters are calculated from the data starting from a time point $T_0$ rather than the initial frame to ensure that the simulation has reached a statistically stationary state.

The global flow order parameter is the displacement of the particles normalised by the displacement of the run without shot, quenched noise and repulsion with the same number of particles, size, and number of frames. Since we are using periodic boundary conditions in most of our simulations, taking the displacement by taking the difference between the initial and the final position would be misguiding so that we sum over the velocities for each agent throughout the analysing time interval and then calculate the displacement from that.\\ 

The calculation could be summarised as follows:

\begin{equation}
Global\:Flow = \frac{1}{\Delta\mathbf{x}_{max}} \frac{1}{N} \sum_{i = 1}^N \|\sum_{T_0}^T {\mathbf{v}_i(t)} dt \| \\   
\label{Global Flow}
\end{equation} \\
where $\Delta\mathbf{x}_{max} =  \|\sum_{T_0}^T {v_{align}} dt \|$ is denoting the maximum displacement in a run without obstacles and repulsion, hence dividing by it will normalise the results. A value close to 1 corresponds to the particles having a relatively large displacement while the value 0 means that there is no displacement on average. This latter case may occur if the density of the obstacles is so large that nearly all particles become trapped/blocked.\\

The average direction is calculated according to:

\begin{equation}
\phi^{ave}=\frac{1}{T-T_0} \frac{1}{N} \sum_{T_0}^T \|\sum_{i=1}^N {\mathbf{\hat{v}}_i(t)} dt \|
\label{Average Direction}
\end{equation}
where $\mathbf{\hat{v}}_i(t)$ denotes the unit vector in the direction of the velocity of particle $i$ at time $t$. This order parameter quantifies how similar the directions of the particles are. A value of 1 denotes a state where all the particles are moving towards the same direction while a value of 0 represents no average direction of the individual motion of the particles. In simple systems the above two order parameters would be nearly the same, however, for complex motion patterns - see later - they can exhibit a characteristic deviation.\\

Throughout our paper the notion "rotating particles" means particles circling (or rotating) around some centre of motion (and not rotating around their own centre of imaginary mass (interaction force). The rotational ratio is obtained from the expression:

\begin{equation}
Rotational\:Ratio = \frac{N_{rot}}{N}
\label{Rotational_ratio}
\end{equation}
where $N_{rot}$ denotes the number of particles which are rotating. Each particle is counted as rotating if its motion in more than $90\%$ percent of the frames, corresponds to a rotational motion. A particle is considered to be in a rotational motion trajectory in a time interval if it is going through a nearly circular path and it is coming back close to its initial position. The details about the algorithm are described in Methods, Quantifying rotations.\\

Lastly, global rotation is calculated as:

\begin{equation}
Global\:Rotation = \frac{|N_{ACW} - N_{CW} |}{N}
\label{Global_rotation}
\end{equation}
where $N_{ACW}$ denotes the number of particles rotating  anticlockwise and $N_{CW}$ denotes the number of particles rotating clockwise. Here a value 1 corresponds to a situation in which  all of the particles are rotating in the same direction whereas 0 means that half of them are rotating CW (ACW) while the other half half rotating ACW (CW), respectively or no particle is rotating at all. For more details, see the Methods section.

\subsection*{States/patterns of collective motion}

Based on our order parameters we define six different collective motion patterns/phases describing the various collective states of our system. These states are summarised in Fig. \ref{motion_states}.  In this section we will go through their definition.\\
(a) \textit{Directional Motion} (Fig. \ref{motion_states}a) is associated with a behaviour when the agents are having a well defined common direction of motion, i.e., they are moving as a coherent flock. This state is represented by  large (close to 1) global flow, large average direction and small rotational ratio and global rotation. For details of the particular threshold values between states look at the  Methods, Motion states section.\\
(b) \textit{The labyrinth state} (Fig. \ref{motion_states}b) is a mixture of states being locally analogous to the directional or the  rotational state. This state contains groups of particles having either directional or rotational motion for a time interval and then changing their motion state. This state has values in middle ranges for all of the order parameters. \\
(c) \textit{Synchronised rotation} (Fig. \ref{motion_states}c) and (d) \textit{Rotation} (Fig. \ref{motion_states}d) are our rotational states. Both have small global flow, mid range average direction and large rotational ratio. The difference between the two rotational states is that while in the synchronised rotation state all of the particles are rotating in the same direction (so that global rotation is large), in the rotation state there are two kind of groups of particles rotating - in a locally synchronized manner -  in different directions (CW or ACW) which leads to smaller global rotation values.\\

The above mentioned four states (a) to (d) are our main states since we can reach all of them only by changing a single parameter, the obstacle density. In addition, we can observe in our simulations two additional states which we call Random (Fig. \ref{motion_states}e) and Meander (Fig. \ref{motion_states}f). The random state is completely random meaning the absence of both directionality or any sort of ordered rotation. Here the particles are going through a fully random motion. The meander state is very much like the meander state in Ref. \cite{Chardac2021}. This state is similar to the labyrinth state by being a mixture of the directional and the rotational states but here the particles are not forming groups of the same behaviour. In addition, temporal analysis supports that the velocity field in the labyrinth state is dynamic whereas the velocity field in meander state is nearly static in analogy with the vortex glass state of Ref. \cite{Chardac2021}.

\section*{Results}

\begin{table}[h!]
\centering
\begin{tabular}{ |p{3cm}||p{3cm}|p{8cm}|  }
 \hline
 \multicolumn{3}{|c|}{Default Value of the Parameters} \\
 \hline
Parameter & Default Value & Definition \\
 \hline
 particle density & 1 & Number of particles divided by area \\
 ts & 1 & Time step\\
  \hline
 $v_{align}$ &   0.02  & Alignment velocity\\
 $r_{align}$ & 2 & Alignment radius\\
 $\eta$ & 0.01 Radian & Shot noise\\
  \hline
  $c_{rep}$ & 0.006  & Repulsion coefficient\\
$c_{core}$& 1  & Hard-core repulsion coefficient\\
 $r_{particle}$    & 0.25 & Hard-core radius of particles\\
 $r_{obstacle}$&   0.25  & Hard-core radius of obstacles\\
 $r_{cut-off}$& 1  & Cut-off radius of soft repulsion\\
A & 0.44 & Controls soft-repulsion threshold\\

\hline

\end{tabular}
\caption{Default value of the parameters.}
\label{table:1}
\end{table}

Most of our simulations were carried out in square box assuming periodic boundary conditions. In order to demonstrate the flexibility of our approach, we also considered the case of a circular box with a wall repelling the particles with a distance dependent force corresponding Eq. \ref{Repulsion}. (For detailed information look at Methods, Boundary Conditions) The initial positions of the particles and the obstacles were drawn from a uniform distribution (thus, their centres were randomly distributed in the simulational area).

We explored a wide range of parameters in the simulations. On the other hand, to concentrate on the most interesting patterns, in the majority of the cases we used a set of default values for almost all of the parameters while varying only the remaining few ones (typically the obstacle density). These default parameters are shown in Table 1. In the following we summarise our main findings concerning the four relevant aspects of the motion patterns appearing in the simulations of our model. 

\begin{figure}
\centering
\includegraphics[scale = 0.80]{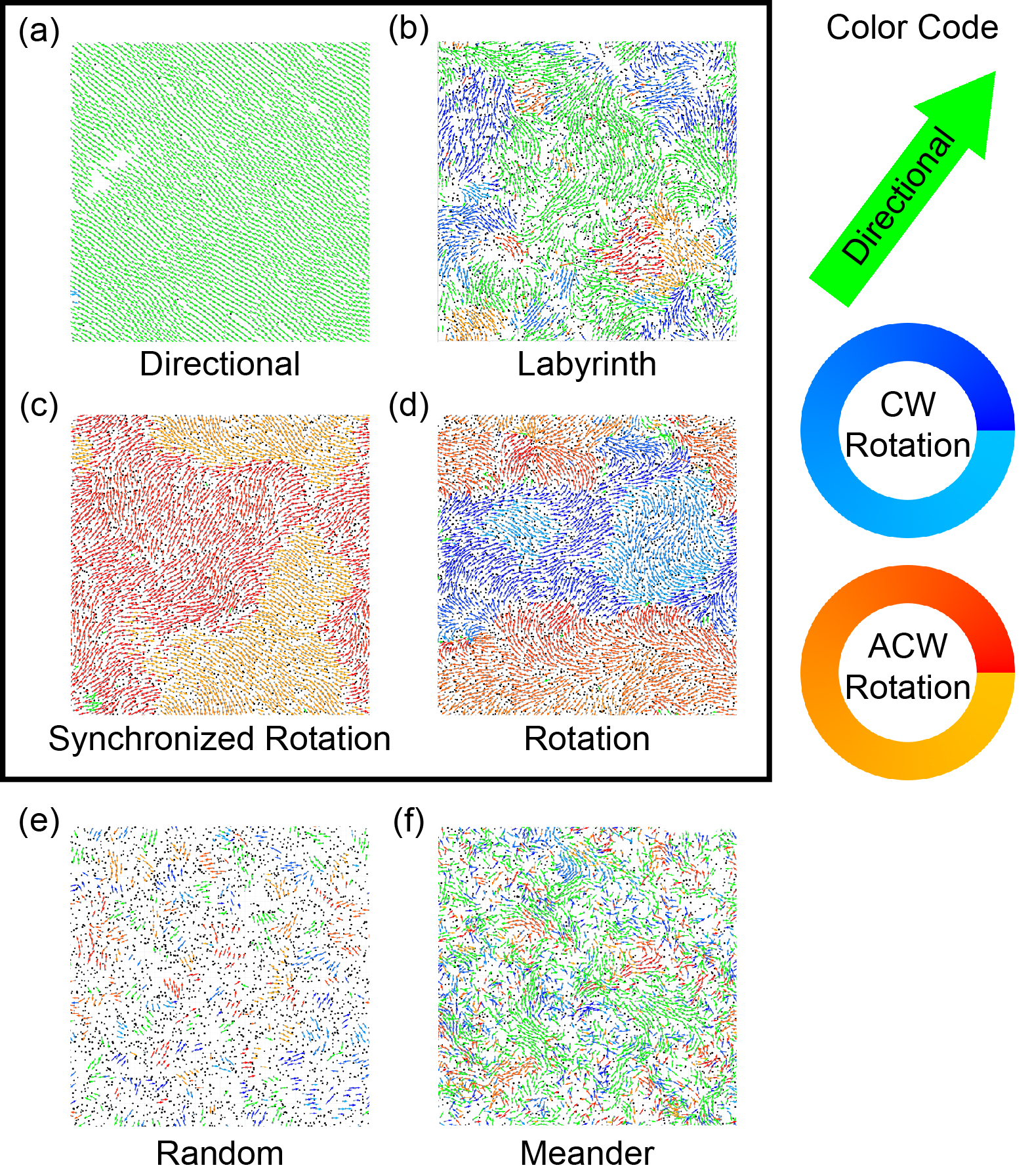}
\caption{Visualization of the motion patterns. All of the figures are for boxes with size 60x60 and using periodic boundary conditions. The figures (a) to (d) were obtained by using the values of the default parameters specified in Table 1. with changing the number of obstacles only. Thus, the simulations were carried out for obstacle densities  0.005, 0.39,0.45 and 0.66, respectively. As for figure (e) we used all of the parameters same as given in Table 1. except that we used particle  density = 0.22 and obstacle density = 0.66, while for figure (f) was obtained for  $r_{align}$ = 0.75, $r_{obstacle}$ = 0.2, $r_{particle}$ = 0.1, $c_{rep}$ = 0.033, $v_{align}$ = 0.2 and obstacle density = 0.22. (a) Particles move as a directional flock, (b) A mixture state of coexisting rotational and directional groups of particles where the spatial boundaries of the two mixture states and the velocity field are changing during the simulations. (c) Synchronised rotation state where all the particles are rotating in the same direction. (d) particles rotating in both directions. (e)  Random state of motion. No directionality, no rotation. (f) A mixture state between the rotational and directional behaviour. Here, in contrast to the labyrinth state the particles move individually and the velocity field is nearly constant. We also visualized these motion states in the videos included into the Supplementary Material (see Supplementary videos 2-8, showing frames at every fifth time step)}. 
\label{motion_states}
\end{figure} 
\bigskip

Our main results about the possible complex collective motion patterns are visualized in Fig. \ref{motion_states} and by the corresponding videos (see Supplementary material). In order to simultaneously represent both the momentary direction of motion and the direction of rotation, we used an innovative colour coding scheme as shown at the right side in Fig. \ref{motion_states}. Particles which do not rotate are indicated by by green arrows, while those rotating clockwise or anticlockwise are shown using shades from blue to turquoise and red to yellow, respectively.

\begin{figure}
\centering
\includegraphics[scale = 0.8]{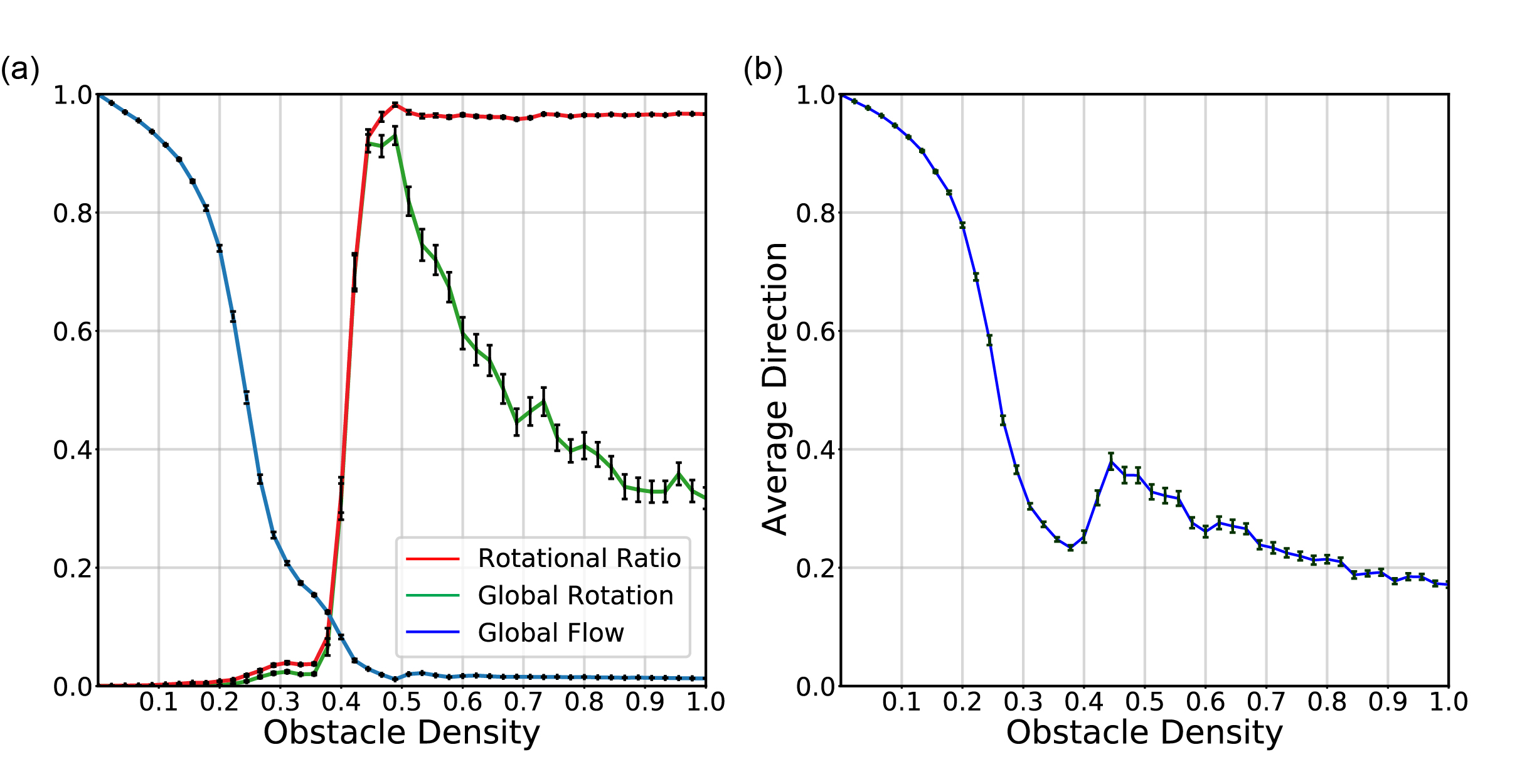}
\caption{Order parameters vs. obstacle density. The figures are drawn for 160 simulations in 60x60 boxes with periodic boundary conditions and parameters being the same as in Table 1. The error bars denote the standard error of the averages. The blue line in (a) indicates the order parameter of the directional motion of state. It seems to be the dominating/only pattern of motion till obstacle density 0.3. In the interval 0.3-0.45 the particles are in a mixed state which we associate with a labyrinth-like pattern. The change around an obstacle density 0.4 signals the emergence of the synchronised rotation state as a growing ratio of the particles are rotating.  The Global Rotation increases and then beyond appr. 0.45 decreases, the initial increase shows the emergence of synchronized rotation state and the decrease shows the transition from synchronised rotation to rotation where both directions exist (with synchronization on a local scale only). The initial decrease of the average direction (b) shows the loss of directionality in the motion of the particles, while the relatively sudden increase at 0.45 is due to the emergence of synchronized rotation.}
\label{order_parameters}
\end{figure} 

An essential aspect of behavioural changes in systems consisting of many similar units is the nature of the level of order and the transitions between the various observed states or, as they are called in equilibrium systems: phases. Although our system is in a far from equilibrium regime, we also consider that phenomena occurring during flocking in most of the cases can be interpreted in terms of approaches common in equilibrium statistical mechanics (See, e.g., \cite{Stumpter2010,Vicsek2012,Marchetti2013,Cavagna2014,Baglietto2009}). This fact was our motivation for defining order parameters and displaying these parameters in Fig. \ref{order_parameters} - as a function of the obstacle density playing the role of the "control parameter" in our case. The changes in the three order parameters depicted in Fig. \ref{order_parameters}(a) indicate behaviours that are novel in several respects. The gradual decay of the global flow parameter is, on the one hand, familiar from earlier studies demonstrating that introducing randomness (both shot noise kind \cite{Vicsek2012} or quenched noise through obstacles - see, e.g., \cite{Chardac2021}) into flocking results in a decay of the level of coherence in the flocking state. 

On the other hand, it should be noted, that in our case there seems to be an anomaly just before the global flow completely ceases. Although, at the first sight, this might be attributed to finite size effect, a further inspection of Fig. \ref{order_parameters} adds a twist to this deviation from a usually monotonic (as far as concerning the first derivative of the plot) decay to zero. In fact, in the region of the obstacle density values between (approximately) 0.3 and 0.45, as the global flow approaches zero, a simultaneous, and very different kind of motion pattern appears, i.e., the global rotation order parameter starts to increase. According to our simulations (which are shown for medium box sizes of 60x60, but we could confirm similar behaviour for larger sizes as well) the motion state that emerges in this region is best described by particles rotating in confined areas "surrounded" by obstacles. The quote mark here stands for a specific and novel aspect of a soft active matter system with obstacles: even if the obstacles are distributed completely randomly the particles - interacting with each other and the obstacles - spontaneously respond to such an environment as if it had "cages" or "chambers" within which they move as if this sort of confinement was analogous to a well defined geometrical area (e.g., circle or square). It is important to point out here that such a behaviour emerges mostly due to the soft nature of the interaction and the relatively large density of both the active particles and the obstacles.

Perhaps the most conspicuous aspect of this emerging rotational state is that the direction of the rotation of the particles is synchronized over the whole system within this (narrow) region of the obstacle density. This is also due to the softness and the relatively large densities of the active particles, so that they can interact even if they are not within the same local cage but are moving in circles in neighbouring ones. This is a delicate state since both the alignment and the repulsion rules act against such a synchronization because of the following reason. Particles rotating in the same direction and getting close to each other in neighbouring cages would move in opposite direction (while this is disfavoured by the alignment rule). The system seems to self-organize into a delicate state in which the particles rotate in neighbouring cages in such a way that most of the time they are as much further away from those rotating in the neighbouring cages as possible. In this way both the cages and the pattern of motion is selected in a self-organized manner so that the state is optimal from the point of minimizing conflicting configurations (collisions, see,\cite{Helbing1999}). The transition from partial to full synchronization accurs gradually, although the slope of the order parameter function is steep. However, due to the corresponding error bars its increase is definitely different from vertical. Fig. \ref{order_parameters}(a) also shows that beyond obstacle densities close to 0.45 although almost all particles keep rotating (the rotational ratio tends to stay close to the value of 1), the level of synchronization (rotating in the same direction throughout the system) gradually decreases and a state made of groups or "patches" of particles rotating in the same direction within a group (but rotating in the same or oppositely directions in different groups) is increasingly taking place.

\begin{figure}
\centering
\includegraphics[scale = 0.2]{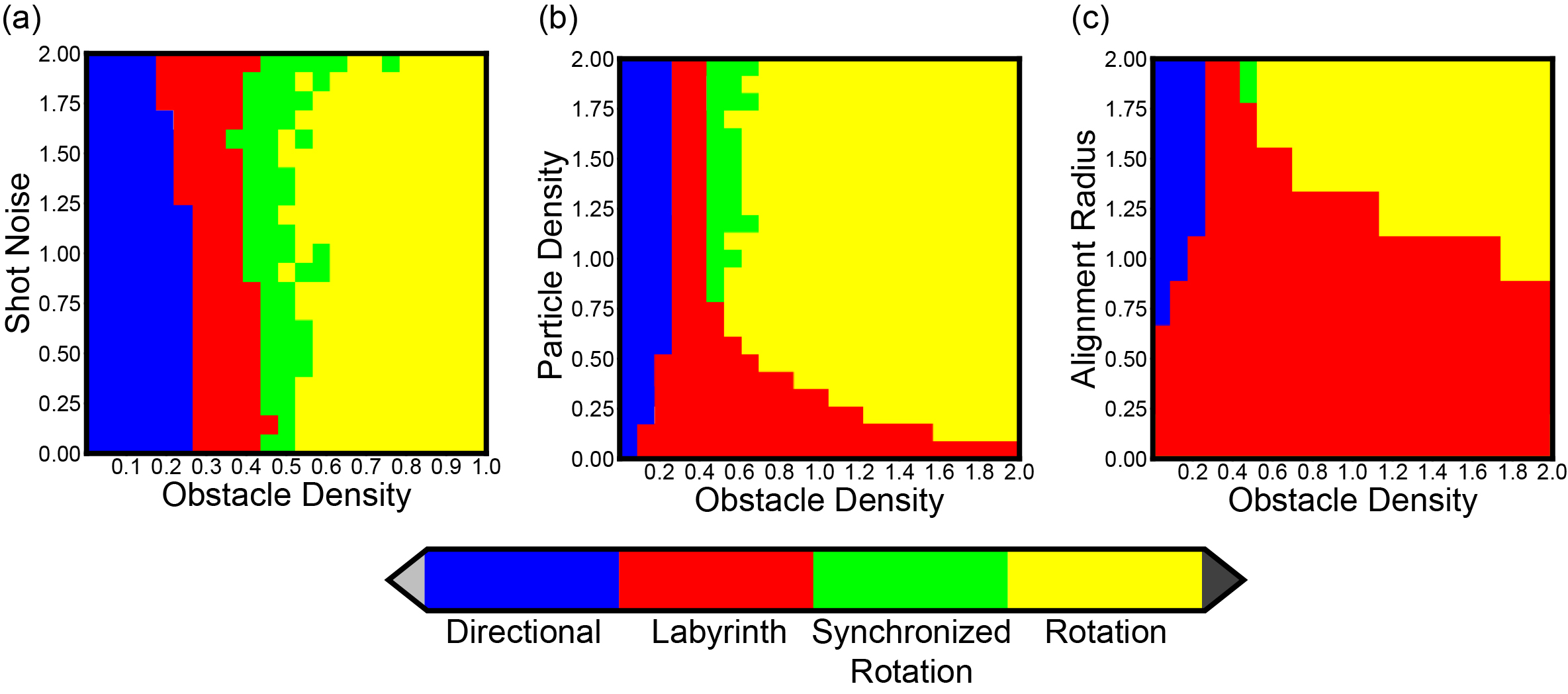}
\caption{Phase Diagrams. The results are shown for 60x60 box size and obtained for periodic boundary conditions. In each panel a)-(c), the parameters, except the parameter along the Y-axis, are the same as those in Table 1. Figure (a) indicates that the obstacle density (quenched noise) dominates shot noise since there is no relevant change in the behaviour over much of the region of the shot noise values. Figure (b) shows the trade between the particle and obstacle densities. It shows that a dense medium is needed to see synchronized rotation and rotation. Figure (c) resembles figure (b) since the alignment radius is related to an effective particle density.}
\label{Phase_diagram}
\end{figure} 

The observation of the above collective motion patterns (phases) calls for exploring the parameter spaces for which they occur. Thus, next we constructed the corresponding phase diagrams in which the regions characterised by a given phase are visualized using four colours (Fig. \ref{Phase_diagram}). Constructing such diagrams needs a considerable amount of CPU/GPU time, this is the reason for some ruggedness in the plots. However, the main tendencies are well seen: the behaviour is considerably less sensitive to the level of shot noise than that of the quenched one (obstacle density), while the way a given motion state depends on the density of the particles or their alignment radius is similar (as expected).

\begin{figure}
\centering
\includegraphics[scale = 0.8]{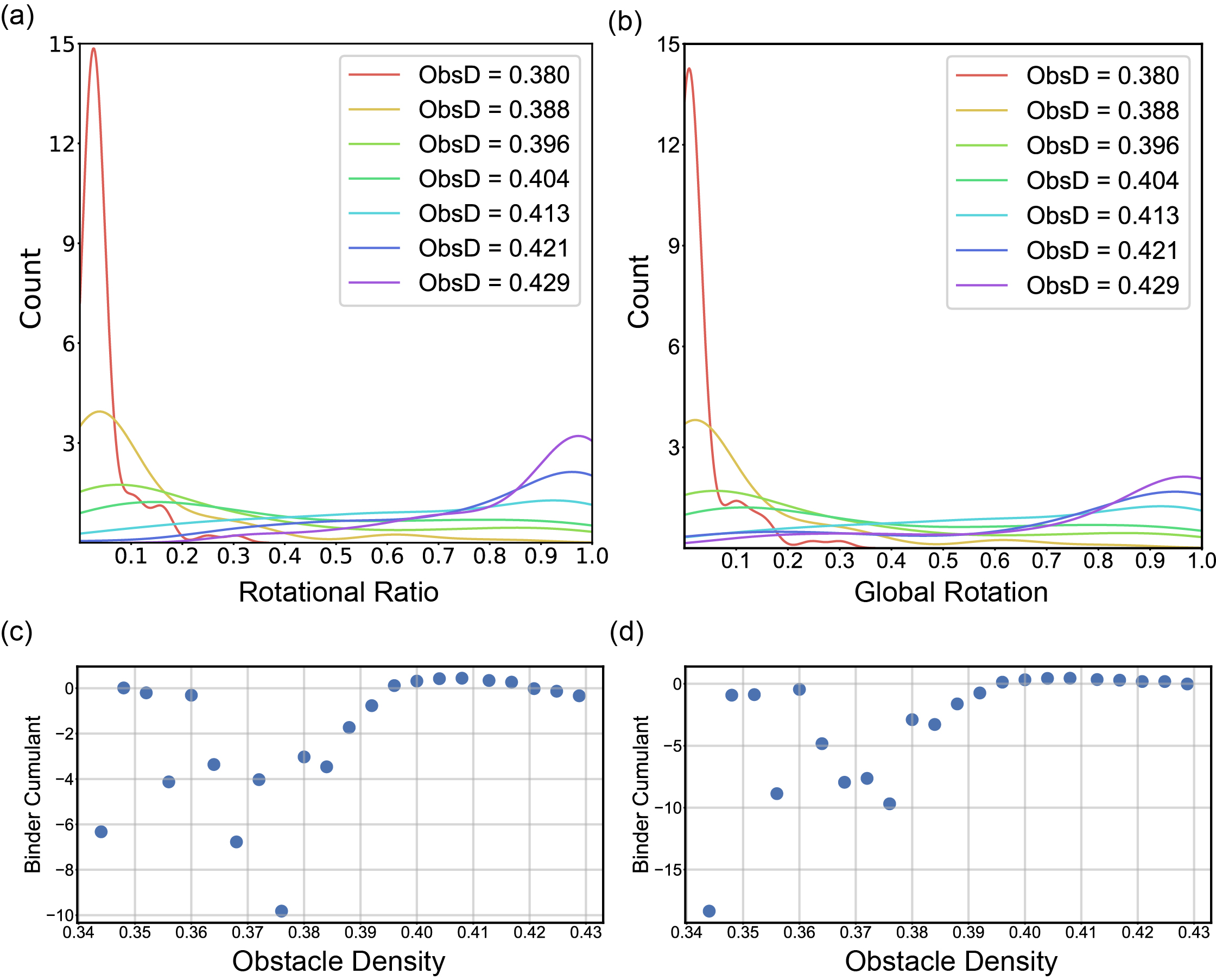}
\caption{Binder cumulant and Histograms. Each data point in (c-d) was obtained from 90 runs in a 100x100 box with parameters same as in Table 1. For interpretation of the above see the text.}
\label{BC_hist}
\end{figure} 

Both Fig. \ref{order_parameters} and Fig. \ref{Phase_diagram} suggest that examining the characteristic feature of the transition from the global flow to the global rotation states is of particular interest. Since the transition seems to be relatively sharp we decided to test whether it is of first or second order. To obtain related information we made use of two approaches (see Fig. \ref{BC_hist}): i) plotting the distribution of the measured global rotation values averaged over many runs and ii) calculating the Binder cumulant \cite{Binder1981}. In case of a standard second order phase transition, the histograms of the order parameter have a well defined maximum gradually moving from smaller to larger values of the position of the maximum as ordering takes place. In the case of a first order transition the behaviour is different, and the position of the maximum moves from the minimum (close to zero) to its other peak without a smooth moving from left to right (decreases for the small values and starts to increase for values close to 1, corresponding to the system assuming either a disordered or an almost fully ordered state. The Binder cumulant signals the order of the transition by either having no minimum at all (second order transition) or having a sharp, well defined minimum at the point of the first order transition. Since both the histograms and the Binder cumulant are sensitive to fluctuations we carried out the corresponding simulations for larger (100 x 100) system/box sizes.

Our Fig. \ref{BC_hist} (c) shows a behaviour that is quite different from any related results reported earlier. We suggest that this unusual statistic of the Binder cumulant is likely to indicate the presence of a transition that does not fall into the category of neither a second order nor a first order transition - a statement to which we shall return briefly in the discussion. A further specific feature of the data plotted in Fig. \ref{BC_hist} is that the anomalous behaviour of the Binder cumulant is more expressed for obstacle densities for which the double peaked nature of the histograms does not show up yet.

\begin{figure}
\centering
\includegraphics[scale = 0.8]{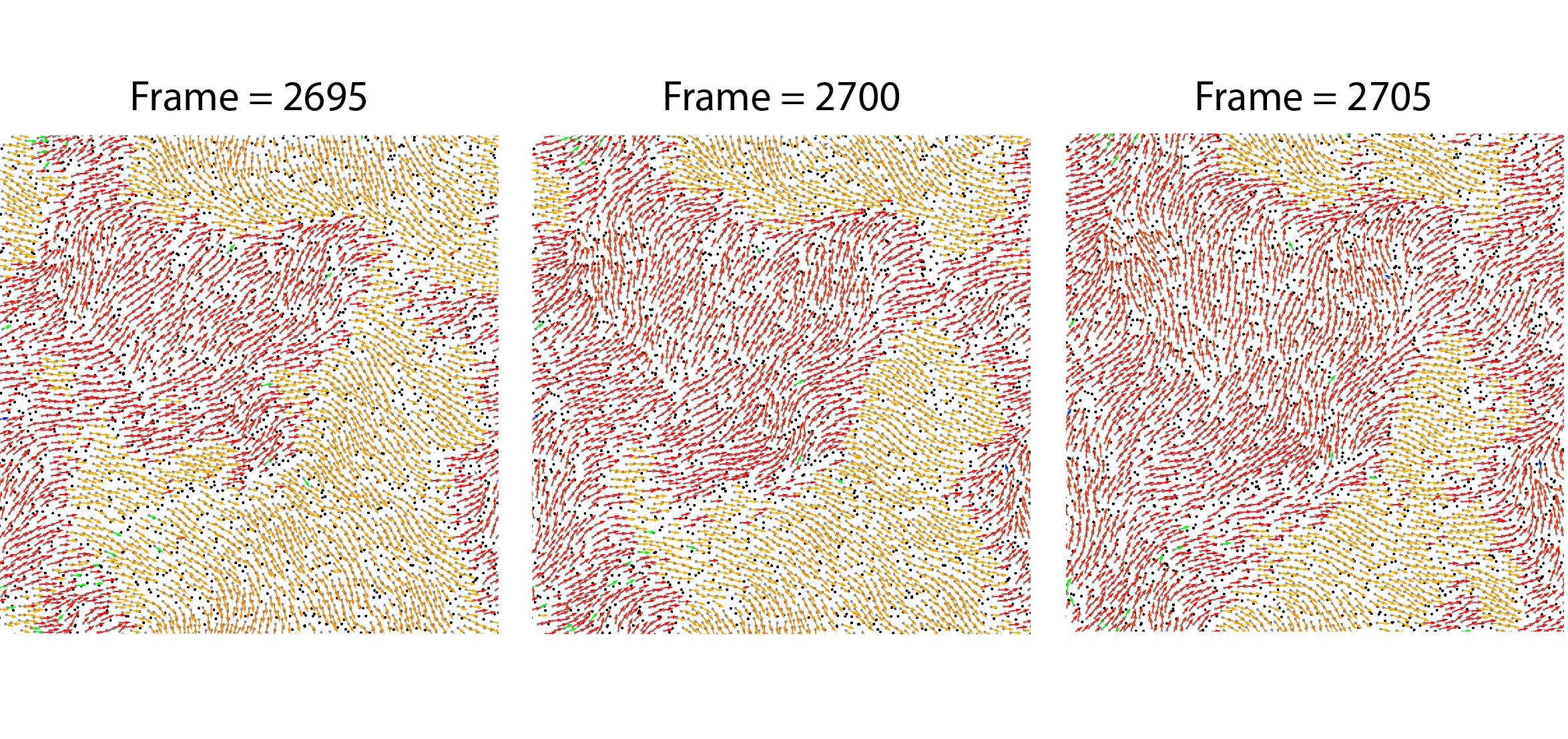}
\caption{Visualization of phase waves. Snapshots from Supplementary video 4 (Synchronised Rotation). Here the word phase is used for denoting the monetary angle of motion of the particles. At each iteration step, each particle transfers its phase to the neighbouring particles which leads to a moving wave-like behaviour. The propagation of the phase wave is indicated by the advancing of the region with red colour, denoting angles close to, but larger than  0 according to the colour code in Fig. \ref{motion_states}.}
\label{Phase_wave}
\end{figure} 

The final point we would like to present is concerned with the way a given direction of the rotating particles propagates in the simulation area. Fig. \ref{Phase_wave} demonstrates this feature: as the number of time steps is increased the borderline between regions of red colour (corresponding to particles moving in a direction above "horizontal") and yellow colour (particles moving in direction below "horizontal") has a tendency to proceed with a velocity depending in a complex way on the local distribution of the obstacles. We include this observation because of its similarity to propagating waves in an excitable medium (for a recent review see Ref. \cite{Zykov2018}). The analogy is based on assuming that the momentary direction of motion corresponds to a given local state of the system and this state propagates across the system in a repeated manner.

\section*{Discussion}

We have demonstrated above that the simple model we introduced in order to account for possible behaviours of dense active matter made of soft particles moving in a disordered medium may give rise to a rich and not yet observed set of collective motion states and the transitions between them. The most unexpected state we have found is made of particles exhibiting two types of synchronization simultaneously. They rotate in the same direction locally, along nearly circular trajectories of average diameter which is of the same order as their interactions radius and, at the same time, they are rotating in the same direction (either all of the clockwise or anticlockwise) at the distant points of the system as well. 

Together with the other outcomes of our simulations our model is unique in the sense that it seems to couple a number of features typical for quite different alternative systems as follows: i) our model is constructed in the spirit of active matter systems or flocking, ii) it displays synchronized rotations occurring in models of synchronization \cite{Strogatz2000} and in a recent model called "swarmalators" \cite{Strogatz2017} that couples swarming to rotations, iii) the nature of the transition between the global flow and global rotation states is likely to be analogous to non-reciprocal phase transitions (Ref. \cite{Fruchart2021}) and iv) we observe the propagation of a state variable just like it happens in excitable media \cite{Zykov2018}. We can add two comments of general nature which are indicated by our simulations: it seems that (at least for some parameter values) the system spontaneously self-organizes into a stationary state in which the number of collisions is minimal and, since our approach allows to control the two kinds of noises independently, we conclude that it is likely that in active matter systems the role of quenched noise dominates over shot noise type perturbations.

\section*{Methods}
\textit{Simulations:}  To efficiently simulate large number of particles, we used a GPU-Accelerated code which is based on CUDA Python. The results were obtained by using  GEFORCE RTX 3060 and GEFORCE GTX 1070 GPU-s. At each time step, the velocity of each particle is calculated in parallel with other particles depending on the current positions and the directions of the particles and obstacles within its repulsion and alignment radius as given by the equations \ref{V_sum} to \ref{Repulsion}. The order parameters were calculated by a combination of CPU and GPU computing to make the calculation more efficient in the steps where paralleling over the particles or time step would not shorten the computational time significantly. \\
\textit{Quantifying Rotations:}
Quantifying the rotational behaviour is essential in our research. To obtain data for interpreting the rotational motion of particles and their full path  with perfect accuracy one needs to store all the positions and the angles and then go through the data. Since this is computationally heavy, we made use of an approximate algorithm: we defined a trajectory as "rotation" if a particle completed a nearly circular path while its velocity vector was going through a full cycle. This corresponds to the following: for each particle and at each iteration, we calculated the rotation of the velocity vector as "rotation" at time $t$:
\begin{equation}
Rotation_{i}(t) = \theta_i(t+1) - \theta_i(t)
\end{equation}
It is important to mention that our angles are stored in the $[0,2\pi]$ interval and we want to know the "direction" of the rotation. The issue of using the mentioned equation is that  it will not give any information of the direction of the rotation and it will only give the angle difference. For example, if the  $\theta_i(t) = 0$ and  $\theta_i(t+1) = \frac{\pi}{2}$ the code correctly calculates the size of rotation as $\pi$, but it will not understand whether the rotation was clockwise or anti-clockwise. To handle this situation, we use the approximation that the velocity vectors cannot have a rotation more than $\pi$ which is reasonable considering our default parameters given in Table. \ref{table:1}. As a result of this approximation, the given example will be calculated as $\pi$ radian rotation in anti-clockwise direction rather than $\frac{3\pi}{2}$ in clockwise direction. Using the same approximation, we also calculated the rotations close to the $0$ or  $2\pi$ point. For example, if $\theta_i(t) = \frac{5\pi}{3}$  and $\theta_i(t+1) = \frac{\pi}{3}$, we have a rotation of  $\frac{2\pi}{3}$ in the anti-clockwise direction rather than $\frac{4\pi}{3}$ in the clockwise direction.\\

After these calculations, we sum over the rotations and the displacement of the particle in a time interval until: Firstly, the rotation sum reaches either -2 $\pi$ or 2 $\pi$ and secondly the displacement becomes smaller than a threshold value which is equal to the hard-core radius of particles ($r_{particle}$) in the current code. If both constraints are satisfied, we consider the particle to be rotating in that time interval and the length of the time interval is counted as the period of the rotation. In addition, if the rotation sum is positive, it is counted to be a positive rotation or ACW and if it is negative, it is a negative rotation or CW.
The first constraint ensures that the path is nearly circular, and the combination of two together ensures that the particles are coming back to their starting positions.
Further, if it takes for a particle longer than a Maximum Possible Rotation Time (MPRT) to satisfy both constraints, then the sums are made equal to zero and we start over. This constraint is extremely important to make the model more accurate. Without this the algorithm would not consider a motion being rotational if a particle goes through a straight path and then starts rotating after a certain time point.
By tuning the threshold and the MPRT value, the algorithm perfectly counts the rotational motions in our simulations since we are only interested in counting full cycles.

\textit{Motion States:}
The following table depicts the threshold values of the order parameters used for defining the various motion states:

\begin{table}
\centering
\begin{tabular}{ |p{5cm}||p{2cm}|p{2cm}|p{2cm}||p{2cm}|  }
 \hline
 \multicolumn{5}{|c|}{Order Parameters} \\
 \hline
Motion state & Global Flow & Average Direction & Rotational Ratio & Global Rotation \\
 \hline
Directional  & $\geq $ 0.5  &  $\geq $  0.5 & $<$0.3 & $<$ 0.7 \\
Labyrinth & $<$ 0.5 &  $<$ 0.5 &   $<$ 0.7 & $<$ 0.7\\
Synchronized Rotation &  $<$ 0.5 & $<$ 0.5 & $\geq $ 0.7 &  $\geq $ 0.8\\
Rotation &  $<$ 0.5 & $<$ 0.5 & $\geq $ 0.7 &  $<$ 0.8\\
 \hline

\end{tabular}
\caption{Order Parameters for each state.}
\label{table:2}
\end{table}
\textit{Boundary Conditions:}
In circular box simulations, we simulate the wall as a chain of obstacles using Eq. \ref{Repulsion} with small modifications:

\begin{equation}
\begin{split}
\mathbf{v}^{wall rep}_i(t+1) = - c_{rep}  \Big(\exp{\frac{r_{i} + r_{rep} - \|R_{circ} - \mathbf{r}_i(t)\|}{A}} - 
\\
- c_{core} g(r_{i} + r_{rep} - \|R_{circ} - \mathbf{r}_i(t)\|) \Big) \mathbf{\hat{r}}_{i}(t)
\end{split}
\label{Repulsion_wall}
\end{equation}
where $R_{circ}$ is the radius of the circular box and $\mathbf{r}_i(t)$ is the radial position vector.
\\

\section{Supplementary Videos}
The supplementary videos are available at \href{https://www.youtube.com/playlist?list=PLnAspyyQS2v1dUveemFM_nQuQlLPNBEx0}{Playlist}

\section{Acknowledgments}
The authors thank S. H. Strogatz for illuminating suggestions and M. Nagy, F. Pakpour and L. L. Sz\'anth\'o for useful discussions. This work was partly supported by the grant: (Hungarian) National Research, Development and Innovation Office grant No K128780. D. V. is grateful to the Erasmus+ Traineeship Programme of the European Commission for support.

\section{Authors' contribution}
T.V. Conceived the research. D.V. performed the calculations. Both authors discussed the results, drew conclusions and edited the manuscript.

\section{Code availability}
The main body of the code used in simulations is available at \href{https://github.com/Danial-Vahabli/Vicsek_Model_with_Repulsion_Cuda}{GitHub} \\
The full version of the code may be available from the corresponding author upon reasonable request.

\section{References}

\bibliographystyle{naturemag.bst}
\bibliography{Citation}

\clearpage
\clearpage
\end{document}